\documentclass[12pt, letterpaper]{article}
\usepackage[utf8]{inputenc}
 \usepackage{graphicx}        % standard LaTeX graphics tool
\usepackage{amstext}
\usepackage{amssymb}
\usepackage{amsmath}
\newtheorem{theorem}{Theorem}
\newtheorem{definition}{Definition}
\newtheorem{lemma}{Lemma}
\newtheorem{corollary}{Corollary}

\begin{document}
 
 \title{Model and Set-Theoretic Aspects of Exotic Smoothness Structures on 
$\mathbb{R}^4$\footnote{To appear in: {\it At  the  Frontiers  of  Spacetime:  Scalar-Tensor Theory, Bell’s Inequality, Mach’s Principle, Exotic Smoothness}, ed. T. Asselmeyer-Maluga (Springer, 2016), in honor of Carl Brans’s 80th birthday.}}
\author{Jerzy Kr\'ol\\University of Silesia, Institute of Physics,\\ ul. 
Uniwersytecka 4, 40-007 Katowice, Poland, \\ iriking@wp.pl}
%\and Name of Second Author \at Name, Address of Institute 
%\email{name@email.address}}
\maketitle

\abstract{Model-theoretic aspects of exotic smoothness were studied long ago 
uncovering unexpected relations to noncommutative spaces and quantum theory. 
Some of these relations were worked out in detail in later work. An important 
point in the argumentation was the forcing construction of Cohen but without a 
direct application to exotic smoothness. In this article we assign the 
set-theoretic forcing on trees to Casson handles and characterize small exotic 
smooth $R^4$ from this point of view. Moreover, we
show how models in some Grothendieck toposes can help describing such 
differential structures in dimension 4. These results can be used to obtain the 
deformation of the algebra of usual complex functions to the noncommutative 
algebra of operators on a Hilbert space. We also discuss the results in the 
context of the Epstein-Glaser renormalization in QFT.}

\section{Infinite geometric constructions and set-theoretic forcing}
\label{sec:0}
Currently it is a bit of a folklore to say that dimension 4 is exceptional both in physics and mathematics. On the one hand this is the 
dimension where Einstein theories of relativity were formulated, where the physics 
of particles and quantum fields found their marvelous realization on (curved) 
Minkowski spacetimes, and where the cosmological evolution of our world is to be 
described. On the other hand, many curious mathematical facts, like the existence of exotic $R^4$, or in fact, of a continuum many of them, take place exactly in this dimension. It was a big effort of many mathematicians in 1980's like Donaldson, Freedman, Gompf, Taubes and many
others whose work on topology and geometry of manifolds in dimension 4 opened our eyes on the unique 4-dimensional topological and `smooth' world and help in its understanding.  However, taking seriously advanced and technical mathematical findings as applicable to physics, required much scientific imagination and courage in those days. It was Carl Brans who took the step in a series of papers \cite{Brans1980,Brans1994,Brans1994a,Brans1999}. Soon after, there appeared the work 
of Torsten Asselmeyer-Maluga (e.g. \cite{Asselm1996}) and Jan S{\l}adkowski (e.g. \cite{Slad1996,Slad2001}) who approached the role of 
exotic $\mathbb{R}^4$'s in physics from various perspectives. Carl's Brans ideas and the papers above were an inspiration to me and I have been lucky as a 
researcher to work together with Torsten and Jan within the recent years. It is a 
big honor and pleasure to me to contribute to the volume celebrating the work of Carl Brans.

Exotic smoothness structures on $\mathbb{R}^4$ are just Riemannian, curved smooth 4-manifolds (exotic $R^4$) which topologically are (homeomorphic to) $\mathbb{R}^4$. In this chapter, I will show that the perspective of set theory and Grothendieck 
toposes, hence foundations of mathematics, is the right one when considering physical applications of exotic, open 4-smoothness. Even though this is neither obvious nor widely accepted approach, the use of model and set-theoretic methods in physics has a firm and vivid tradition arisen from the foundations of mathematics (e.g. \cite{Lawvere1975,Benioff1976,Takeuti1978,Kock1981}). That was developed substantially further in recent years (e.g. \cite{Benioff2002,Benioff2015,Isham2007,Isham2010,Landsman2007,Krol2004}).

In physics, set theory is usually considered informally as unchanged eternal background which goes together with the classical 2-valued logic. However, when one allows for variations in such background more formal, axiomatic formulation is needed. That is why set theory is understood as the first order axiomatic Zermelo-Fraenkel (ZF) theory of sets with possible addition of the axiom of choice (AC) -- ZFC. Similarly arithmetic is usually described as an axiomatic first order theory - Peano arithmetic (PA) (see the discussion regarding the order of formal theories vs. set theory in \cite{Vaananen2001}). The variations in the theories can be grasped by considering various models of these theories. Classically such models (Tarski) are built in the category Set of sets and functions between them. All models of first order theories undergo usual limitations and benefits which follow the Goedel or Loewenheim-Skolem-like theorems (and much more, see  e.g. \cite{Jech2003}).  
We also will be using more general models of (intuitionistic) set theory in other categories like toposes where the logic becomes intuitionistic \cite{Moerdijk1992}. 

The forcing method is known from the independence results in set theory since 1960's \cite{Cohen1963} and allows for changing the models. In general, forcing in mathematics is a very rich, technical and advanced subject (see e.g. \cite{Jech2003,Bart1995}). For the purpose of this work it is a method for studying the real numbers line.  
Thus Cohen forcing in a narrow sense used in the chapter can be seen as a mechanism of adding real numbers to the model and thus changing the model of ZF(C) and the real line. This is also a tool for exploring the exotic smooth $R^4$'s (see e.g. \cite{Krol2004,Krol2004b,Krol2003}). 

We start with infinities appearing in some geometric constructions in dimensions 3 and 4 like Casson handles and Alexander's horned sphere (wild embeddings). These infinities are the inevitable and intrinsic features of the constructions. On the other hand, infinity by itself is a natural and central topic in set theory. The key for understanding this relation is precisely the Cohen forcing. On the algebraic level a forcing is generated by some complete atomless Boolean algebra - in this case the forcing is nontrivial and can eventually add some reals to the ground model $M$ of ZFC. In the case of Cohen forcing the algebra is the \emph{unique} atomless Boolean algebra with a dense countable subset. In fact it holds true:
\begin{lemma}[Corollary 25.4, p. 189 \cite{JustWeese1997}]\label{-1}
Let $A$ be a complete atomless Boolean algebra that contains a countable dense 
subset. Then $A$ is isomorphic to the algebra ${\rm RO}(CS)$ of regular open 
subsets of the Cantor set CS.
\end{lemma}
Any (signed) tree canonically generates a partial order (partially ordered set). A partial order $(\mathbb{P}, \leq)$ is called \emph{separative} if for all $p,q\in \mathbb{P}$ such that $p \nleq q$ there exists $r\leq p$ with $r \perp q$. Here $r \perp q$ means incompatibility relation i.e. there does not exist $k$ that neither $q\leq k$ nor $r\leq k$ is true. Then, it holds true the important lemma:
\begin{lemma}[Lemma 13.33, \cite{JustWeese1997}]\label{-3}
Every separative partial order $\mathbb{P}$ can be completed to a complete 
Boolean algebra $B$ such that $\mathbb{P}$ is dense in $B\setminus\{0\}$ and the 
partial order in $\mathbb{P}$ agrees with $\leq_B$. $B$ is unique up to 
isomorphism.
\end{lemma}
Next we ask the question: which rooted trees do represent a separative partial order? One easily finds that the full binary tree (the one which has precisely 2 branches at every node) does. Moreover:
\begin{lemma}\label{-2}
The full binary tree represents the countable dense subset (partial order) of 
some complete atomless Boolean algebra.
\end{lemma}
This is because the full binary tree represents the Cantor set in $(0,1)$ interval: one assigns to every branch $0$ or $2$ numbers which appear in the three-mal decompositions $0.x_1x_2x_3...$ of numbers in $(0,1)$. Then missing numbers correspond precisely to $x_i=1, i=1,2,3,...$. The nodes of the tree represent the members of the countable partial order which is dense in the partial order of the tree hence in the corresponding Boolean algebra. The algebra is ${\rm RO}(CS)$ which is atomless and generates the nontrivial Cohen forcing.  $\square$

Now the point is that the Cantor set generated by the binary tree is frequently realized geometrically by Casson handles construction in dimension 4 and by wild embeddings of spheres in dimension 3 (see e.g. \cite{Freedman2013}).
Casson handles (CH) (see e.g. \cite{F1982,GS1999,TAMCB2007}) appear in the handle-body decompositions of small exotic smooth open 4-manifolds \cite{GS1999} are also represented by the infinite signed rooted trees \cite{GS1999,Bizaca1995}.  If the tree was finite and the CH smooth, the Casson handle would be the ordinary smooth 2-handle\footnote{Every CH is topologically (as a pair) homeomorphic to the standard 2-handle which was shown by Freedman \cite{F1982}.}.

Let me quote an important and elementary observation by Kato (\cite{Kato2004}, p. 114) which ensures that given a signed tree we have a Casson handle spanned on that tree:
\begin{quotation}
There are sufficiently many Casson handles. In fact to each infinite signed 
tree, one can associate a Casson handle.
\end{quotation}
Let $M$ be a model of ZFC and $M[G]$ its generic extension by Cohen forcing \cite{Jech2003,Bart1995}. Then we can prove the following:
\begin{theorem}
A general Casson handle appearing in the handlebody of a small exotic $R^4$ 
determines a nontrivial Cohen forcing adding a Cohen real in some generic model 
$M[G]$ of ZFC.
\end{theorem}
\emph{proof}:
 First, any Casson handle can be embedded in the simplest CH which is the linear 
tree with only one, positive or negative, self-intersection at each level. This 
follows from the fact that every CH with a bigger signed tree than the tree of 
another CH is embeddable in this `smaller-tree-CH'. One should respect the rule 
that the smaller tree is homeomorphically embedded into the bigger one. Adding 
self-intersections on any level and killing the generators by gluing kinky 
handles determines the embedding. Moreover, the resulting embeddings of CH's 
preserves the attaching areas of CH (or at least attaching circles and their 
framings). The last means that whenever the simplest CH were exotic (the 
attaching circle determines the non-smooth slice) the embedded CH with a bigger 
tree would be exotic too \cite{Bizaca1995}. 

Second, instead of attaching an arbitrary CH let us attach the simplest one (see figs. \ref{simplestCH} and \ref{exotic r4}) with the linear signed tree in which we know the bigger one is embeddable. In 
general we do not know whether the CH with such a tree is exotic although we know 
it is exotic for the `only +' or `only -' trees.

Next, let us consider the Casson handle determined by the full binary tree 
(BT) with one infinite branch identical with the linear one above. Such 
`binary-tree-CH' embeds in the linear CH and let us forget the signs in the 
binary CH. Then from Lemmas \ref{-2} and \ref{-1} the algebra ${\rm RO}(CS)$ is 
the unique Cohen forcing algebra generated by BT.  
$\square$

Note that every CH determines the same (up to isomorphism) Cohen algebra thus 
the nontrivial Cohen forcing in a generic model $M[G]$. In dimension 3 given wildly embedded 3-sphere, say horned Alexander sphere, a `grope' is assigned naturally to it which is spanned on the infinite binary tree again (\cite{Freedman2013}, pp. 18-19). Thus Cohen forcing can be built also in this case. We do not discuss the meaning of it here but note only that wild embeddings in dimension 4 are second sides of exotic open 4-smoothness and this can be understood physically as a quantum state \cite{TAMJK2011,TAMJK2011b}.  

Cohen forcing changes the real line substantially, namely the reals in the model $M$ constitute merely measure zero subset of the extended real line in $M[G]$, hence of $\mathbb{R}$. As shown above it is also assigned to replacing the standard smooth 2-handles by an exotic Casson handle, hence to changing the smoothness structures on $\mathbb{R}^4$. If the forcing acted over $\mathbb{R}$-line in $\mathbb{R}^4$ and resulted in exotic $R^4$ the following important question would arise: Can an extension of the real line by forcing be a valid tool when exploring exotic smoothness in dimension 4? In some sense this kind of forcing should add reals to the full $\mathbb{R}$ resulting in the same $\mathbb{R}$ since $R^4$ is again the Riemannian smooth real manifold. We will analyze this problem in the next and subsequent sections.

\section{From the standard to categorical $\mathbb{R}^4$}
\label{sec:1.2}
One needs 'adding' more real numbers to the already full $\mathbb{R}$. What is the meaning of such procedure? We will show that the modification of logic and set theory is needed. 

From the external absolute point of view a set-theoretic forcing adds 
reals (if any at all) to subsets $R_M$ of $\mathbb{R}$ where $R_M$ is a set of real numbers in 
some model $M$ of ZFC. Internally there is no difference between (1st order) 
properties of real lines $R_M$ and $\mathbb{R}$. Suppose that we already have a well defined model of the standard real line $\mathbb{R}$.\footnote{A formal 
theory giving rise to the unique up to isomorphism model of real numbers should use the 2nd order logic. Such theories are called categorical (in 
$\aleph_1$). The theory of Archimedean complete ordered field is categorical. It 
is a second order theory.} 
Starting with $\mathbb{R}$ can one add consistently more reals to the line? More 
precisely: can one construct a bigger real line which would have \emph{the same} 
properties as $\mathbb{R}$ but be different as a set (thus containing more reals)? Our general motivation for considering such questions, as observed in the 1st section, is that we expect 
such procedure to possibly modify the smoothness of manifolds.  

Reducing the properties of the real line to its 1st order properties, and the 
logic to first order logic, Robinson showed \cite{Robinson1966} that there are non-standard models of arithmetic ${}^{\ast}N$ and analysis ${}^{\ast}R$. They are end-extensions of 
the standard $\mathbb{N}$ and $\mathbb{R}$ respectively and contain infinite 
natural and real numbers. Moreover, ${}^{\ast}R$ contains infinitesimal 
invertible real elements. Now, every true 1st order formula $\phi$ about natural 
numbers is fulfilled in ${ }^{\ast}N$ iff it is fulfilled in $\mathbb{N}$, i.e. 
${ }^{\ast}N\vDash \phi\,\equiv \, \mathbb{N}\vDash \phi$. We say that ${ 
}^{\ast}N$ and $\mathbb{N}$ are elementary equivalent and write: 
\begin{equation}\label{E1}
{ }^{\ast}N \simeq_{1} \mathbb{N}\;\; ({ }^{\ast}R \simeq_{1} \mathbb{R}),
\end{equation}  
meaning, one can not distinguish the two models just by their 1st 
order properties. We would like to strengthen the indistinguishability as above 
and consider something like ${ }^{\ast}N \simeq_{2,3,...} \mathbb{N}\;\; ({ 
}^{\ast}R \simeq_{2,3,...} \mathbb{R})$.\footnote{It would be sufficient to 
consider ${ }^{\ast}N \simeq_{2} \mathbb{N}$ since there are theorems reducing 
the higher order to 2nd order logic (e.g. \cite{Hintikka1955,Montogue1965}).} It is seemingly a trivial task, 
since 2nd order theory of natural or real numbers are categorical and the real 
line $\mathbb{R}$ is the only (up to isomorphism) model allowed, hence indeed ${ 
}^{\ast}N \simeq_{2,3,...} \mathbb{N}$. 

That is why we are rather looking for an environment (the twist) where 
non-standard models for arithmetic and analysis may exist, are nontrivial, i.e. 
different, and are valid for higher order theories, i.e. some second order 
properties of the models become identical after the twisting. Without any twist 
these particular properties would not coincide. As noted above we can not 
achieve the nontrivial realization of the full classical indistinguishability ${ 
}^{\ast}N \simeq_{2} \mathbb{N}$ $({ }^{\ast}R \simeq_{2} \mathbb{R})$ since 2nd 
order arithmetic has isomorphic models. 

To imagine how the twist could work one can introduce three parameters $(w,\alpha,\epsilon)$ controlling the twist - $w$ corresponds to the 
weakening of the arithmetic and/or the logic, and the other two to the fractions (belonging to $(-1,1)$) of the numbers of all true formulas of the first and second orders correspondingly. $\alpha=0$ and ${w}=0$ mean 
that all true 1st order formulas of both models, $({ }^{\ast}R$ and $\mathbb{R})$, are determined with respect to 
the first order (i.e. $\alpha=0$) classical (i.e. $w=0$) predicate logic. Similarly, $\epsilon=0$ and $w=0$ mean that all second order formulas of the models are determined w.r.t. the 
classical second order logic. Thus one writes 
\begin{equation}\label{E2}
{ }^{\ast}N \simeq_{1-\alpha,1+\epsilon}^{w} \mathbb{N}\; ({ }^{\ast}R 
\simeq_{1-\alpha,1+\epsilon}^{w} \mathbb{R})
\end{equation}
when the logic is weakened and the sets of the first order formulas and second 
order formulas have been modified and especially some 2nd order formulas 
become identical in both models after the twist. The $+,-$ signs indicate the twist or the rotation in the parameter space. 
The value of the parameters depends on the degree of how much of 
weak and nonclassical logic is used. We do not need to determine the relation between the parameters more precisely here.
Instead, let us consider the important example. We will weaken the logic and arithmetic 
considerably and take the models in a constructive set-up, i.e. in toposes. 

This weak Peano arithmetic was recognized in detail by Moerdijk and Reyes \cite{MR1991} when they considered the non-standard models of numbers in smooth toposes and build the smooth topos model for 
synthetic differential geometry. We present the discussion of the elements of their construction important for us in the Appendix \ref{App}. 

The important point is that the objects of natural numbers (NNO) in smooth toposes like Zariski (${\cal Z}$) and Basel topos (${\cal B}$) determined by the natural embedding of manifolds from Set to the toposes, i.e. the map $s:\mathbb{M}\to {\cal Z}$, sends $\mathbb{N}$ to the standard natural numbers $s(\mathbb{N})$ in ${\cal Z}$, ${\cal B}$\footnote{As the object in a topos this standard NNO is the constant sheaf of natural numbers.}, fails to generate a proper object of real numbers $s(\mathbb{R})=R_{\cal Z}$ (or $R_{\cal B}$). For example: $R_{\cal Z}$ is nonarchimedean with respect to 
$s(\mathbb{N})$ so thus (\ref{5}) does not hold. Besides $[0,1]\subset R_{\cal 
Z}$ is noncompact with respect to $s(\mathbb{N})$. As the consequence this last 
property devastates the homology theory of manifolds in ${\cal Z}$ (\cite{MR1991}, pp. 280-284.). 

To cure this one should turn to the modified object of natural numbers 
$N_{\cal Z}$ (smooth natural numbers) which is not the canonical standard NNO 
$s(\mathbb{N})$ in ${\cal Z}$. As shown by Moerdijk and Reyes the axioms of the weak logic 
(\ref{1},\ref{2},\ref{3}) are fulfilled in ${\cal Z}$ however the type $N$ is 
interpreted now as $N_{\cal Z}$ i.e. it is the smooth NNO. $R_{\cal Z}$ is now Archimedean w.r.t. 
$N_{\cal Z}$, $[0,1]$ is compact (smooth compact, or $s$-compact), the 
homologies of manifolds are tractable and in particular the internal topologies 
of manifolds in ${\cal Z}$ are well-defined. Internal in ${\cal Z}$ constructions and 
theories are formulated such as the true natural numbers are $N_{\cal Z}$ rather than 
the standard $s(\mathbb{N})$. The shift $s(\mathbb{N})\to N_{\cal B}$ changes some second order properties of real and natural numbers such that now in ${\cal Z}$ internal constructions are more like the external ones.

The construction of $N_{\cal Z}$ follows the filterproduct construction. Namely, 
the object $R_{\cal Z}\simeq {\rm s}(\mathbb{R})$:
\[ R_{\cal Z}=s(\mathbb{R})=\mathbb{L}(-,lC^{\infty}(\mathbb{R})) \] is the 
representable object of ${\cal Z}$ \cite{MR1991}. It is non-archimedean with respect to 
$s(\mathbb{N})$ as said above. Instead one defines the object of smooth natural numbers 
$N_{\cal Z}$ thus allowing for the modification of \emph{finiteness}. Let 
$(\sin(\pi x))$ be the ideal in $C^{\infty}(\mathbb{R})$. The representable 
object in ${\cal Z}$ of smooth integer numbers $Z_{\cal Z}$ is now defined as (\cite{MR1991}, p. 252)\footnote{$l(\,)$ is the member of $\mathbb{L}$ -- the category of loci which is opposite to the category of (finitely generated) smooth rings (\cite{MR1991}, p. 58).}:
\begin{equation}\label{NNO1}
Z_{\cal Z}=l(C^{\infty}(\mathbb{R})/(\sin{\pi x})), \; N_{\cal 
Z}=l(C^{\infty}(\mathbb{R})/(\sin{\pi x}, x\geq 0)).
\end{equation} 
Taking the ideal $F$ of functions which are non-zero only on finite initial segments of $\mathbb{N}$, then the quotient $l(C^{\infty}(\mathbb{N})/F)$ represents a non-standard infinite natural number in ${\cal Z}$.

To have the standard $s(\mathbb{N})\simeq \mathbb{N}$ one can define it as the 
subtype of $N_Z$:
\begin{equation}\label{7}\mathbb{N}=\{ n\in N_Z:\forall_{S\in P(N_Z)}(0\in 
S\wedge \forall_{n\in N_Z}(m\in S\to m+1 \in S)\to n\in S) \} \end{equation}
which means $\mathbb{N}$ fulfills the strong induction scheme we know from Peano 
arithmetic \cite{MR1991}.  However, when logic is weakened (in the metatheory) the 
'true' natural numbers are defined with respect to  the coherent induction 
scheme (\ref{1}) in which case one does not distinguish $\mathbb{N}$ and $N_Z$. 
We do not dwell upon such metatheoretic considerations here (see however 
\cite{Krol2004}).

Even if the subtype $\mathbb{N}\subset N_Z$ can be defined as in (\ref{7}) still it 
is undecidable:\footnote{A subset $A\subset B$ is decidable when $a\in A$ is 
decidable property, i.e. when $\forall_{a\in B}(a\in A \vee a\notin A)$.}
\[ {\cal Z}\models (\mathbb{N}\neq N_{\cal Z}) \to (\mathbb{N}\; {\rm is\, not\, 
decidable\, in}\; N_{\cal Z}). \]
The important question is the extend up to which one can consistently replace 
$\mathbb{N}$ by $N_Z$. What is crucial here is that the 2nd order property of 
$R_{\cal Z}$ of being Archimedean is again retrieved with respect to $N_{\cal 
Z}$. Similarly, the interval $[0,1]$ is compact again with respect to $N_{\cal 
Z}$. The twist (\ref{E2}) is realized by the shift:
\begin{equation}\label{T1} s(\mathbb{N})\to N_{\cal Z}
\end{equation}
which allows for the retrieving of some internal higher order properties of theories in 
${\cal Z}$ which were lost when the canonical standard NNO was in use. 

We will demonstrate how this intuitionistic model for weak arithmetic and 
especially the shift (\ref{T1}) is related to both smoothness structures in 
dimension 4 and the procedure of adding reals by forcing.

\subsection{Smooth natural numbers in ${\cal B}$}
\label{sec:1.2.1}
Weak logic as described in the previous section (and in the Appendix) guarantees that there is a NNO
different than $s(\mathbb{N})$, i.e. $N_{\cal Z}$ which replaces 
consistently the standard NNO in the intuitionistic set-up. The crucial point is 
that $N_{\cal Z}$ contains also non-standard natural numbers what indicates 
that $N_{\cal Z}$ is an intuitionistic analogue of $ ^{\ast}N$ known from the 
non-standard analysis (NA). Internal in the toposes, higher order intuitionistic 
theories are formulated internally in ${\cal Z}$ w.r.t. $N_{\cal Z}$ and 
$R_{\cal Z}$ leaving aside their standard counterparts. But such radical departure 
from standardness modifies finiteness such that infinite big non-standard natural 
numbers are considered as $s$-finite. 

In general there are two kinds of infinitesimal elements in $R_{\cal Z}$: 
invertible ($\mathbb{I}\subset R_{\cal Z}$) and nilpotent ones. Nilpotent 
elements are required by the synthetic differential geometry approach and they 
represent forms like ${\rm d}x$ (${\rm d}^2=0$), while invertible elements are predicted 
by the non-standard analysis of Abraham Robinson which can be generated by taking 
inverses of infinite non-standard natural numbers. The smooth topos unifies 
both kinds of infinitesimals in the one real line $R$ where they exist as real numbers. 
Moreover, most internal higher order theories perceive the smooth numbers as 
true real and natural numbers. The important class of such theories are 
differentiable manifolds whose category $\mathbb{M}$ is mapped into the smooth 
toposes via $s$ transform, and they require $s$-numbers to define their 
topology, compactness, connectedness or homologies. 

However, do there really exist `non-standard' and invertible 
infinitesimal elements of $R_Z$, i.e. $\mathbb{I}$ in ${\cal Z}$? In fact it 
holds \cite{MR1991}: 
\begin{equation}
Z\models \neg \neg [\exists_x x\in R_Z\cap \mathbb{I}] 
\end{equation}
which is a rather weak version of the existence of invertible infinitesimals 
(recall that the logic in ${\cal Z}$ is intuitionistic and double negation does not 
cancel in general). To strengthen this result the Authors of \cite{MR1991} 
proposed to modify the topos ${\cal Z}$ towards ${\cal B}$ such that now one proves:  
\begin{equation}
{\cal B}\models \exists_x x\in R_Z\cap \mathbb{I}. 
\end{equation}
To obtain this result one has to modify the Grothendieck topology in ${\cal Z}$ and then to be sure invertible infinitesimals do exist, one adds them by the \emph{forcing on stages} (see the Appendix 1 in \cite{MR1991} ). Thus, indeed in the internal environment of ${\cal B}$ the non-standard real numbers are added by forcing. This is the extension of the real line by adding new reals which we discussed in Secs. \ref{sec:0} and \ref{sec:1.2}. Such procedure is not in general possible in higher 
orders and in the classical $\{0,1\}$ logic, but it is possible in the 
weaker logic of the topos ${\cal B}$ realizing the twist (\ref{E2}) by the shift (\ref{T1}).

\subsection{The smooth topos ${\cal B}$ localized on $\mathbb{R}^n$}
\label{sec:1.2.2}
Here we want to show that smoothness structures on $\mathbb{R}^4$ can have their origins at the level of models of the real line. Moreover, continuum many different exotic smoothness structures $R^4$'s can be understood at that level.
Given the real line (higher order, classical) 
$\mathbb{R}$ it is Archimedean with respect to $\mathbb{N}$. To have such a unique 
model $\mathbb{R}$ we can think of it as the model for the second order theory of 
real numbers or the theory of an Archimedean complete ordered field, both having 
unique (up to isomorphisms) models. On the contrary, reducing the properties of $\mathbb{N}$ or $\mathbb{R}$ to the first order we get a plurality of non-standard 
models $ ^{\ast}N$ and $ ^{\ast}R$ in every infinite cardinality. Can one have 
different non-standard models ${ ^{\ast}R}$ all having the cardinality of 
continuum? The answer is the following:
\begin{lemma}
Under the Continuum Hypothesis (or under $2^{<\mathfrak{c}}=\mathfrak{c}$) there 
are $2^{\mathfrak{c}}$ different non-isomorphic models ${ ^{\ast}R}$ all having 
the cardinality $\mathfrak{c}$.
\end{lemma} 
The part of the proof important to us is the observation that every 
non-principal ultrafilter ${\cal U}$ on the set $\mathbb{N}$ generates a 
non-standard $ ^{\ast}R_{\cal U}$ of the cardinality continuum as an ultrapower 
construction, and two such ultrapowers are isomorphic if and only if the 
ultrafilters generating them are isomorphic w.r.t. a permutation of 
$\mathbb{N}$. Finally there are $2^{\mathfrak{c}}$ non-isomorphic ultrafilters 
on $\mathbb{N}$. $\square$

Thus starting with the higher order $\mathbb{R}$ one has up to $2^{\mathfrak{c}}$ 
possibilities to choose its 1st order continuous reducts $\mathbb{R}\to{ 
^{\ast}R}$. This extends to the relation basic to us (especially for 
$n=4$) with 1 to $2^{\mathfrak{c}}$ possibilities:  
\begin{equation}\label{13}
\mathbb{R}^n\overset{2nd\to 1st}{\longrightarrow}{ ^{\ast}R}^n.
\end{equation}
Let us complete this correspondence with another one as follows:
\begin{equation}\label{14}
\mathbb{R}^n\overset{2nd\to 1st}{\longrightarrow}{ ^{\ast}R_1}^n\overset{sh}{\to 
}{ ^{\ast}R_2}^n  \overset{2nd\to 1st}{\longleftarrow}\mathbb{R}^n.
\end{equation}
We would like to have (\ref{14}) realized as smooth correspondence also in the 
middle arrow, and valid in the higher orders. This is the point where the topos 
${\cal B}$ and the twist (\ref{T1}) come into play. We are further extending the 
correspondence (\ref{14}) into the following ${\cal{B}}$-modified one:
\begin{equation}\label{15}
\mathbb{R}^n\overset{2nd\to 1st}{\longrightarrow}{ 
^{\ast}R_1}^n\overset{e_1}{\to} R_{\cal B}^n\overset{[d]}{\to} R_{\cal B}^n 
\overset{e_2}{\leftarrow} { ^{\ast}R_2}^n  \overset{2nd\to 
1st}{\longleftarrow}\mathbb{R}^n.
\end{equation}
We are going to determine the internal in ${\cal B}$ $[d]$-continuous and even 
differentiable map.
Let $\rm Fin$ be the ideal in $P(\mathbb{N})$ of finite subsets of 
$\mathbb{N}$. 
The algebra $P(\mathbb{N})/{\rm Fin}=P(\omega)/{\rm Fin}$ is an atomless Boolean 
algebra. Moreover, all nonprincipal ultrafilters on $\mathbb{N}$ are the members 
of the Stone space $\beta[\omega]\setminus \omega$ of the algebra 
$P(\omega)/{\rm Fin}$. Recall that the Frechet cofinite filter ${\cal F}$ on 
$\mathbb{N}$ is defined as:
\begin{equation}\label{16}
{\cal F}=\{F\in P(\mathbb{N}): \mathbb{N}\setminus F\in {\rm Fin}.  \}
\end{equation}
The following obvious but important lemma holds true:
\begin{lemma}\label{l2}
Every nonprincipal ultrafilter ${\cal U}$ on $\mathbb{N}$ contains the Frechet 
cofinite filter ${\cal F}$.
\end{lemma}
Let us consider now the specific relation of non-standard models $ ^{\ast}N$, $ 
^{\ast}R$ in classical logic (Set) and in toposes (higher order intuitionistic 
logic). 
\begin{lemma}\label{l3}
In ${\cal B}$ and ${\cal Z}$ the non-standard models are built as filterproduct 
constructions based on the Frechet filter ${\cal F}$ rather than on 
ultrafilters.
\end{lemma}
This follows from the direct construction of smooth natural numbers in ${\cal 
B}$ (see \cite{MR1991}, p 252). Moreover, to respect constructivism 
in toposes one cannot base on the AC especially using ultrafilters strongly 
depends on AC. In \cite{Moerdijk1995} Moerdijk showed explicitly that the 
constructive non-standard PA in the topos $Sh(\mathbb{F})$ of sheaves on the 
category of filters is based on the smooth natural numbers constructed with 
respect to the Frechet filter ${\cal F}$.

\begin{corollary}\label{c1}
All non-standard models $ ^{\ast}N$ ($ ^{\ast}R$) are mapped by $e_1,e_2$ in 
(\ref{15}),  into the single intuitionistic non-standard model $N_{\cal B}$ 
($R_{\cal B}$) in ${\cal B}$. 
\end{corollary}
This is the consequence of: (1) All ultrafilters are the extensions of the 
unique Frechet filter (Lemma \ref{l2}).
(2) Different nostandard models of $R$ (with the cardinality continuum) are 
constructed on the base of non-isomorphic nonprincipal ultrafilters on 
$\mathbb{N}$. (3) Lemma \ref{l3}.
 $\square$

Let us consider relations on $\mathbb{N}$ modulo the ideal of finite subsets 
${\rm Fin}$, e.g. the equality becomes $A=^{\ast}B$ meaning $A\Delta 
B=A\setminus B\cup B\setminus A \in {\rm Fin}$.
We call a $1 : 1$ function $f: D_f\to {\rm Im}_f,D_f, {\rm Im}_f\subset 
\mathbb{N}$ an almost permutation of $\mathbb{N}$ whenever domain of $f$, $D_f$, 
and its image ${\rm Im}_f$ are almost $\mathbb{N}$, i.e. $D_f=^{\ast} \mathbb{N} 
= ^{\ast}{\rm Im}_f$.

Each such almost permutation $f$ of $\mathbb{N}$ gives rise to the automorphism 
$d_f$ of the Boolean algebra $P(\omega)/{\rm Fin}$. Namely
\begin{equation}\label{17}
d_f([A])=[f(A\cap D_f)] {\rm \;for\;} [A]\in P(\omega)/{\rm Fin}.
\end{equation}
Even though there can be up to $2^{\mathfrak{c}}$ nontrivial automorphisms of 
the algebra $P(\omega)/{\rm Fin}$ \cite{Walker1974}, it is still valid that:
\begin{lemma}\label{lcont}
There are $\mathfrak{c}$ automorphisms of $P(\omega)/{\rm Fin}$ which give rise 
to almost permutations of $\mathbb{N}$.
\end{lemma}
This is crucial for us to consider such trivial automorphisms since they forbid 
$\mathbb{N}$, hence $\mathbb{R}$, to be constant and give definite 
transformations of $\mathbb{N}$. Moreover, as shown by Shelah \cite{Shelah1982}, the 
statement that there are only $\mathfrak{c}$ automorphisms of $P(\omega)/{\rm 
Fin}$ (only trivial) is consistent with ZF. So in the above sense we restrict our considerations to the 
trivial automorphisms case. Let us note that:
\begin{lemma}
Every trivial automorphism of $P(\omega)/{\rm Fin}$ represented by a permutation $\sigma: \omega \to \omega$ 
corresponds to a mapping (shift) between non-isomorphic non-standard models of 
$\mathbb{R}$ of the cardinality $\mathfrak{c}$. 
\end{lemma}
This is a direct consequence of the relation of the nonprincipal ultrafilters 
and non-standard models of $\mathbb{R}$, and the fact that the Stone space of 
$P(\omega)/{\rm Fin}$, i.e. $\beta[\omega]$, contains all 
nonprincipal ultrafilters on $\omega$, i.e. $\beta[\omega]\setminus \omega$. Every permutation of $\mathbb{N}$ extends to a homeomorphism $\beta(\sigma):\beta[\omega]\to \beta[\omega]$ and to an automorphism of $\beta[\omega]\setminus \omega$ (e.g. 3.41, p. 88 in \cite{Walker1974}). This last defines the shift between the non-standard models. $\square$

In fact we need the following converse relation:
\begin{corollary}
For every automorphism of $P(\omega)/{\rm Fin}$ there exists the shift-map between non-isomorphic non-standard $\mathfrak{c}$-models of 
$\mathbb{R}$ such that the automorphism realizes this shift between the models.
\end{corollary}
Now given the shift-map $sh:{ ^{\ast}R}_1\to { ^{\ast}R}_2$ as in (\ref{14}) we 
can think of it as determined by some automorphism of $P(\omega)/{\rm Fin}$. 
Note that this correspondence is obviously non-unique. Taking an internal in 
${\cal B}$ extension $[d]$ of the shift as in (\ref{15}) gives rise to the 
following:
\begin{theorem}
Every external shift $sh:{ ^{\ast}R}_1\to { ^{\ast}R}_2$ determines the internal 
$s$-differentiable maps $[d]_{1,2},[d]_{2,1}: R^n_{\cal B}\to R^n_{\cal B}, 
n=1,2,3,...$.
\end{theorem}
Note that $[d]_{1,2}$ and $[d]_{2,1}$ are generated in Set by the `inverse' almost 
permutations of $\mathbb{N}$.
\emph{proof}:
First, any non-standard model ${ ^{\ast}R}_i$ is obtained via the ultraproduct 
construction w.r.t. an ultrafilter ${\cal U}_i$. ${\cal U}$ is the extension of 
the Frechet filter ${\cal F}$. In ${\cal B}$ the 'non-standard' real line is 
$R_{\cal B}$ obtained via the \emph{filter construction} w.r.t. ${\cal F}$. Hence we 
have $[d]_{1,2}:R_{\cal B}\to R_{\cal B}$. Second, every internal $[d]$ is 
continuous in ${\cal B}$ (see Theorem 3.6, p. 270 in \cite{MR1991}). Next, since 
${\cal B}$ is the model of synthetic differential geometry (there exist 
indempotant infinitesimals $D\subset R_{\cal B}$) it holds true the Kock-Lawvere 
axiom in ${\cal B}$, which gives (\cite{MR1991}, p. 302):
\[ \forall_{f\in R^R}\forall_{x\in R}\exists!_{f'(x)\in R}\forall_{h\in D} 
f(x+h)=f(x)+h f'(x)  \] where $R$ stands for $R_{\cal B}$.
Note that $f'(x)$ is just the symbol for the unique $y=f'(x)$ such that $y\in 
R$. Repeating the procedure we determine subsequently $f''(x), f'''(x),...$.
Thus $f\in R^R$ is a standardly infinitely many times differentiable internal 
function.
Finally, we apply again the Kock-Lawvere axiom to the 'inverse' map $[d]_{2,1}$ 
which leads to a similar differentiability. 
$\square$
\begin{definition}
The pair $([d]^n_{1,2},[d]^n_{2,1})$, or $[d]^n_{1,2}$ to shorten, is called an 
\emph{internal diffeomorphism} or \emph{$s$-diffeomorphisms} of $R_{\cal B}^n, 
n=1,2,3$.
\end{definition}
Note that any internal diffeomorphism as above is generated by the shift between 
the non-standard $\mathfrak{c}$-models of $\mathbb{R}$. One could wonder whether 
the $s$-diffeomorphisms can be non-identity maps since they all are generated 
w.r.t. the Frechet filter. However, due to Lemma \ref{lcont} there is precisely 
$\mathfrak{c}$ shifts which guarantee that $\mathbb{N}$ hence 
$R_{\cal B}$ are not constant.

Now we are ready to define the central object in this section (cf. \cite{Krol2006}):
\begin{definition}\label{defB}
Let $M^n$ be a smooth $n$-dimensional manifold and $\{U_{\alpha}: U_{\alpha}\in 
{\cal O}\}$ its regular open cover. We call $ ^{({\cal B})}M^n$ \emph{a 
$n$-dimensional manifold $M^n$ locally modified by the topos ${\cal B}$}, or the 
\emph{smooth ${\cal B}$ structure} on $M^n$, whenever it holds:
\begin{itemize}
\item For every regular open cover $\{U_{\alpha}\}$ of $M^n$ there exists some 
$U_{\alpha}\in \{U_{\alpha}\}$ such that $U_{\alpha}$ is internal object of the 
internal in ${\cal B}$ topology of $s(M^n)$.
\item If two such open $U_{\alpha}, U_{\beta}$ are internal in ${\cal B}$ their 
nonempty internal meet defines the local change of coordinates in ${\cal B}$ 
which contain the $s$-diffeomorphisms: 
$\eta_{\alpha\beta}=[d]_{1,2}:U_{\alpha}\cap U_{\beta}\to U_{\alpha}\cap 
U_{\beta}$.
\end{itemize}
\end{definition}
Next we would like to ensure that $s$-diffeomorphisms do not arise from a 
Set-based diffeomorphism. To this end let the class of 
trivial automorphisms of $P(\omega)/{\rm Fin}$ be suitably limited:
one allows only those trivial automorphisms whose almost permutations of 
$\mathbb{N}$ contain at least one non-identical almost cycle - \emph{cyclic almost 
permutations} and we will call them cyclic permutations if it does not cause any confusion.\footnote{An almost cyclic permutation is an almost permutation $p$, i.e. $p:A\overset{1: 1}{\longrightarrow} B, A=^{\ast} \mathbb{N} =^{\ast}B$, which reverses the order of elements of some $C\subset A$ when compared to the order of $p(C)$.} There still exist continuum many such almost permutations and none of them is extendable in Set to any orientation preserving diffeomorphism of $\mathbb{R}$.

Summarizing:
\begin{enumerate}
\item $s$-diffeomorphisms are not images of diffeomorphisms from Set, hence the 
local modification by ${\cal B}$ of the smoothness structure of $M^n$ is 
nontrivial and categorical.
\item $s$-diffeomorphism is generated in Set by a cyclic almost permutation of 
$\mathbb{N}\subset \mathbb{R}$ so it is not extendable to any 
orientation-preserving diffeomorphism of $\mathbb{R}$.
\item\label{i3} In ${\cal B}$ each permutation of $s(\mathbb{N})\subset R_{\cal 
B}$ gives rise to the $s$-diffeomorphism $([d]_{ij},[d]_{ji})$.
\end{enumerate}

\section{From the categorical to exotic $\mathbb{R}^4$}
\label{sec:1.3}
Given the local ${\cal B}$-modification of the smooth structure on $M^n$ we are 
interested in its impact on the actual classical smoothness of manifolds. One 
obvious classical limit (this which does not depend on ${\cal B}$) of the ${\cal B}$-modified structure on $M^n$ is just 
the smooth structure of $M^n$ we started with. In this case all open $U_{\alpha}\in 
{\cal O}$ become (again) Set based external objects. There is however another, 
more refined possibility. 
\begin{definition}\label{defLB}
\begin{enumerate}
\item We say that a classical limit of the ${\cal B}$-deformed smooth structure 
on $M^n$ \emph{factors through the non-standard models ${ ^{\ast}R}_1$ and ${ 
^{\ast}R}_2$} whenever they are $\mathfrak{c}$-models of $\mathbb{R}$ and the 
${\cal B}$-deformation was performed according to (\ref{14}) and (\ref{15}) 
where now ${\cal O}\ni U_{\alpha}\simeq \mathbb{R}^n$ on the l.h.s. and ${\cal 
O}\ni U_{\beta}\simeq \mathbb{R}^n$ on the r.h.s. of these relations. 
\item A \emph{nontrivial classical limit} of the ${\cal B}$-deformed smooth 
structure of $M^n$ is a smooth structure on $M^n$ which factors through some 
non-standard models of $\mathbb{R}$ while reaching the Set and higher order 
levels.
\end{enumerate}
\end{definition}
The point is that even though local ${\cal B}$-modifications of $M^n$ take all 
almost permutations of $\mathbb{N}$ into internal $s$-diffeomorphisms, hence a 
single $\cal B$-deformed structure emerges, on the Set level it is not so. 
Namely we can prove the important result:
\begin{theorem}\label{th2}
For different non-isomorphic $\mathfrak{c}$-models ${ ^{\ast}R}_1$ and ${ 
^{\ast}R}_2$ with the cyclic automorphism shifting them, the classical 
nontrivial limit (if it exists!) of the local ${\cal B}$-deformed structure of 
$\mathbb{R}^4$ is some exotic smooth $R^4_{1,2}$.
\end{theorem}
To fix the result we need the simple but crucial observation:
\begin{lemma}
Given a smooth structure on $\mathbb{R}^4$ if there does not exist any open 
cover of $\mathbb{R}^4$ containing a single coordinate patch $\mathbb{R}^4$ this 
structure has to be exotic.
\end{lemma}
If a smooth $\mathbb{R}^4$ has a single coordinate patch $U\simeq \mathbb{R}^4$ 
it is diffeomorphic to the standard $\mathbb{R}^4$. If none of its open covers 
contains a single element, such $R^4$ can not be diffeomorphic to the standard 
$\mathbb{R}^4$. $\square$
\emph{proof}:(Theorem \ref{th2})
We will show that any coordinate patch of the ${\cal B}$-modified $\mathbb{R}^4$ 
can not contain the single chart. On the contrary let there exist a single 
coordinate patch $\mathbb{R}^4$ for the classical limit of the ${\cal 
B}$-modified $\mathbb{R}^4$ as above. But in this case any open cover can be 
deformed by diffeomorphisms to a cover whose transition functions are 
identities. However, the factorization of some $U_{\alpha}$ through ${ 
^{\ast}R}_1$ and $U_{\beta}$ through ${ ^{\ast}R}_2$ and the cyclic condition on 
the permutations of $\mathbb{N}$ excludes the identities. 
$\square$

Note that this proof works in the case of ${\cal B}$-modified $\mathbb{R}^n$ 
since in this case for the standard $\mathbb{R}^n$ one can have a single 
coordinate patch. It is known that exotic $R^n$'s exist only in dimension $n=4$ 
so that means that classical smooth limits of the categorical ${\cal 
B}$-modifications do not exist for $n\neq 4$. What is so special in dimension 4 
that enables the existence of the limit as above? Some explanation comes from the 
special relation of Casson handles and geometric constructions in dimension 4 
with the smooth NNO in ${\cal B}$.

\subsection{Casson handles in ${\cal B}$}
When proving that emerged smooth $R^4$ is exotic we left aside the case when 
there are external diffeomorphisms which can be mapped onto the internal ones. 
The reason is that they could be `gauged out' to the identity on the 
intersections by some external diffeomorphisms. However, making the additional 
assumption, which is also partly and implicitly present in so far analysis, we 
can include some external diffeomorphisms as generating exotic smooth $R^4$'s. 
Namely, assume explicitly that \emph{natural numbers $N$ are generated as 
different objects by non-isomorphic $\mathfrak{c}$-models of $\mathbb{R}$.} This 
means that given the almost permutations of $\mathbb{N}$ generated by different 
$\mathfrak{c}$-models of $\mathbb{R}$ they can not be `gauged out' to the 
identity whenever the models of $\mathbb{R}$ are non-isomorphic. This is rather 
strong low level assumption  which reverses our `natural thinking' about the 
relation of real and natural numbers. However, in ${\cal B}$ we had a 
similar situation: given the canonical object of real numbers $R_{\cal B}$ which 
is the image $s(\mathbb{R})$ from Set, we had to modify the NNO 
$s(\mathbb{N})$ to the smooth $N_{\cal B}$. The real numbers determined the NNO. 
Now we want to follow this line of reasoning and show that Casson handles are 
related with the smooth NNO in ${\cal B}$. 

\begin{figure}[b]
\centering
\includegraphics[width=\textwidth]{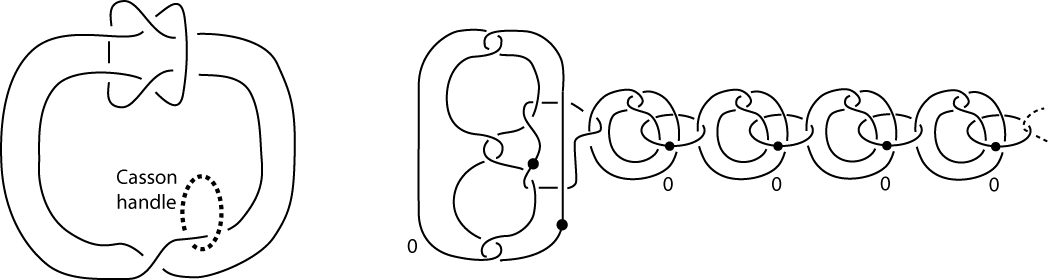}
\caption{The simplest small exotic $R^4$ with the simplest possible Casson handle attached to the Akbulut cork.}
\label{exotic r4}       % Give a unique label
\end{figure}

\begin{figure}[b]
\centering
\includegraphics[width=\textwidth]{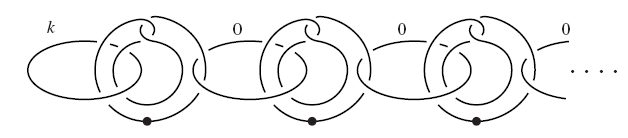}
\caption{The simplest possible Casson handle which gives rise to an exotic 
$R^4$.}
\label{simplestCH}       % Give a unique label
\end{figure}

Let us consider one example.
The simplest known exotic $\mathbb{R}^4$ can be represented in the Kirby 
calculus language as a handle-body with a single Casson handle (fig. 
\ref{exotic r4}, \cite{GS1999}, p. 363). 
The simplest possible Casson handle with a single positive intersection 
at each level (fig. \ref{simplestCH}, see \cite{GS1999}, p. 363).
 Let us assign a partial non-cyclic permutation $p$ of $\mathbb{N}$ to this CH, 
namely define it by: the number of level, i.e. $n$, plus 'the number of 
intersections at each level, i.e. 1. It results in the following partial 
permutation:
\begin{equation}
p: n\to n+1, n\in \mathbb{N}.
\end{equation}
Such a permutation defines the automorphism of $P(\omega)/{\rm Fin}$ according 
to $\ref{17}$ and thus corresponds to the shifts between the 
$\mathfrak{c}$-models of $\mathbb{R}$. Based on the assumption we indeed arrive 
at the exotic $\mathbb{R}^4$.

This simple example justifies the assumption as a basic rule in the context 
of exotic smooth structures on $\mathbb{R}^4$. It also shows that Casson handles 
are nontrivially related with the object of NN in ${\cal B}$. 
One can make this relation even more direct by interpreting the trees spanning 
CH's as built w.r.t. the smooth rather than standard NN. In this case we say 
that a Casson handle is spanned by a tree in ${\cal B}$. Let us turn again to the 
simplest CH represented by its Kirby diagram in the fig. \ref{simplestCH} (see \cite{GS1999}, p. 363). The 
tree is just infinite $+$-signed linear order of levels. The crucial information 
is its infiniteness resulting from the geometric construction. More precisely:
\begin{lemma}\label{l7}
If the smooth Casson handle construction terminated after finitely many steps it 
is the standard smooth 2-handle.
\end{lemma}
This means that any smooth $\mathbb{R}^4$ with a handle-body containing 
\emph{all} smooth finite CH's becomes the standard smooth $\mathbb{R}^4$.
Let us now associate the smooth NN to the levels of the simplest CH:
\[ \# \; {\rm of\;level} \to n\in \mathbb{N}\subset N_{\cal B} \] just by taking 
the infinite set of levels as complementary to the finite set $\{0 \}$ thus 
becoming a member of the Frechet filter ${\cal F}$. But this means that the 
infinite tree of this CH is just $s$-finite in ${\cal B}$. When one performs 
similar enumerating of infinite number of levels in an arbitrary CH the result 
is the following:
\begin{lemma}\label{l8}
Infinite Casson handles are spanned in ${\cal B}$ by $s$-finite trees.
\end{lemma}
This together with Lemma \ref{l7} indicates that indeed the internal arithmetic 
of $\cal B$ has something to do with exotic smoothness, since one can state:
\begin{corollary}
Exotic smoothness structures on $\mathbb{R}^4$ (smooth 4-manifolds), while 
transformed into ${\cal B}$ by $s$, belong to the class of $s$-standard smooth 
$\mathbb{R}^4$. They all are internally $s$-diffeomorphic.
\end{corollary}  
This result in fact agrees with our previous observation that external 
distinct, even discontinuous maps lead to internal $s$-diffeomorphisms. What was crucial in establishing it was the shift (replacement) from the standard 
$\mathbb{N}$ to the smooth $N_{\cal B}$. The same shift is crucial in the above 
seeing CH's as $s$-finite objects.
Observe that turning to the locally modified by ${\cal B}$ structures of 
manifolds, allows for the shifting between various exotic $\mathbb{R}^4$'s, not 
necessarily between exotic and the standard ones. Namely it holds:
\begin{theorem}
Let $R^4$ be a small exotic $\mathbb{R}^4$ whose handle-body contains $k$ many 
CH's for some $k\in \mathbb{N}$. Let a local ${\cal B}$-modification of $R^4$ be 
performed such that $l<k, l\in \mathbb{N}$ $l$-many CH's belong to the local 
open neighborhood which is internal in ${\cal B}$. Then, there exists a 
classical limit of this modification which is an exotic $R^4_{k-l}$ (with only 
$k-l$ nontrivial CH's).
\end{theorem}
\emph{proof}:
Observe that internally $l$ CH's becomes $s$-finite CH's (those corresponding to 
the $s$-finite spanning trees). It is enough to define the classical limit as 
$R^4_{k-l}$ by requiring that $s$-finite CH's are sent to the actually finite 
ones. 
$\square$

Now, we see that the local modification of the manifold smooth structures by 
${\cal B}$ and taking classical limits, works as an analog of the large 
diffeomorphism where the actual smooth exotic $R^4$'s represent a 
kind of \emph{generalized} isotopy classes of embeddings (or small, 
coordinate-like diffeomorphisms). Working entirely in Set one can not realize 
exotic $R^4$'s as merely isotopy classes of embeddings since there is no 
diffeomorphism at all connecting different exotic $R^4$'s. Moreover, in this 
generalized set-up, one can study a class of topological and smooth manifolds 
allowing for the local categorical modifications (and the resulting new concept 
of equivalence). The local character of the modification leads to 
\emph{generalized manifolds} which are partially both in Set and ${\cal B}$. 

\section{Some consequences to Physics}
\label{sec:1.4}
Starting with $\mathbb{R}$ and ${ ^{\ast}R}$ and creating the pairs of such 
reals for both models we arrive at the isomorphic fields of complex numbers, 
even though $\mathbb{R}$ and ${ ^{\ast}R}$ are non-isomorphic. 
This is connected with the fact that in $\mathbb{C}$ one can not define the NNO 
$\mathbb{N}$ (starting from the axioms of the complete ordered algebraically 
closed field of characteristic zero). 
But this means that we can use ${ ^{\ast}R}$ instead of $\mathbb{R}$ in the case 
of the complete ordered algebraically closed field of characteristic zero, i.e. 
$\mathbb{C}$.

Given the divergent expression $1+2+3+4+...=\sum_{i=1}^{\infty}i$ it is bigger 
than any $n\in \mathbb{N}$ so this sum, if existed as the natural number (and in 
1st order language), corresponds to a non-standard number of some model ${ 
^{\ast}N}$ hence ${ ^{\ast}R}$. Moreover, such non-standard element exists in any 
non-standard $\mathfrak{c}$-model of $\mathbb{R}$, since every non-standard model 
of $\mathbb{N}$ is the (conservative) end-extension of $\mathbb{N}$. 

Note that we get the same $\mathbb{C}$ (up to isomorphism) starting from any ${ 
^{\ast}R}$ by building the space of pairs with the algebraic operations of 
$\mathbb{C}$. This is the consequence of categoricity of $\mathbb{C}$.
Thus, possibly the non-standard big values, like the infinite sum above, should 
correspond (via the isomorphisms of models) to some finite value in 
$\mathbb{C}$. 

Indeed, suppose such value does not exist, then each pair of the form 
$(\sum_{i=1}^{\infty}i,b), b\in { ^{\ast}R}$ can not correspond to any standard 
complex number. But it does since every ${ ^{\ast}C}\overset{\rm iso}\simeq 
\mathbb{C}$ ($\mathbb{C}$ is $\mathfrak{c}$-categorical). 
Moreover it has to correspond via the isomorphism to some standard pair $z\in 
\mathbb{C}, z=(x,y); x,y\in \mathbb{R}$. The point is the following: 
$\mathbb{C}$ allows $2^{\mathfrak{c}}$ nontrivial automorphisms and they give 
rise to the isomorphisms ${ ^{\ast}C}\overset{\rm iso}\simeq \mathbb{C}$ for 
every ${ ^{\ast}C}$ generated via the ultrafilter constructions. On the other 
hand there are only 2 automorphisms of $\mathbb{C}$ that send $\mathbb{R}$ to 
$\mathbb{R}$ - the identity and the complex conjugation. This, together with the 
fact that fixed points of all automorphisms of $\mathbb{C}$ are all rational 
numbers, i.e. $\forall_{\phi \in {\rm Aut}(\mathbb{C})}\forall_{r\in 
\mathbb{Q}}\phi(r)=r$, give that the image of $\sum_{i=1}^{\infty}i$ under any 
isomorphism ${ ^{\ast}C}\overset{\rm iso}\simeq \mathbb{C}$ has to be irrational 
pair $(x,y)\in \mathbb{C}: x,y\in \mathbb{I}$. This is in fact result of a very 
discontinuous and wild behavior of the (wild) automorphisms of $\mathbb{C}$ 
realizing the above isomorphism. On the other hand if one would like to have a 
finite value assigned to this iso which would not be dependent on the choice of 
the non-standard model ${ ^{\ast}R}$ it has to be rational number as it is a 
fixed point of every automorphism. 
In what follows we would like to consider this model-theoretic mechanism for 
assigning finite values to divergent expressions in context of exotic smoothness 
structures on $\mathbb{R}^4$. Then we try to understand this phenomenon in 
context of renormalization and regularization ever-present in perturbative 
quantum field theories.\footnote{This part of the work was performed in the 
cooperation with Krzysztof Bielas.} 

\begin{lemma}
For any exotic smooth $R^4$ (which is topologically $\mathbb{R}^4$) any 
diffeomorphic image of it can not send smooth coordinate line $\mathbb{R}$ to 
the smooth $\mathbb{R}$.
\end{lemma}
If there were such diffeomorphism the exotic $R^4$ would factorize as 
$\mathbb{R}\times \mathbb{R}^3$ which is necessary standard. $\square$

One can equivalently state the lemma as: \emph{If the topological $\mathbb{R}$ 
is smooth line in a smooth $R^4$ this has to be standard $\mathbb{R}^4$.} Thus, 
when a smooth diffeomorphism of $R^4$ preserves $\mathbb{R}$ as the factor this 
can happen only for the standard $\mathbb{R}^4$.  
In the case of automorphisms of $\mathbb{C}$ when $\mathbb{R}$ is send to 
$\mathbb{R}$ then the automorphism can not be wild. Otherwise, any wild 
automorphism scatters in a very discontinuous way the real line in the complex plane 
(leaving the rational numbers fixed). For any exotic diffeomorphism of $R^4$ it 
can not smoothly send the line $\mathbb{R}$ to itself, though continuously it does. As we explained in the previous sections and in this one, both 
situations are connected with non-standard $\mathfrak{c}$-models of 
$\mathbb{R}$. 

Let us consider the non-standard ${ ^{\ast}C}$ (though isomorphic to 
$\mathbb{C}$) as generated by pairs of the non-standard reals, i.e. ${ 
^{\ast}C}\simeq \{(a,b)\in { ^{\ast}R}\times { ^{\ast}R}$\}. Then, make the 
product: ${ ^{\ast}C}^2\simeq { ^{\ast}R}^4$. When turning to the higher orders 
one gets the unique (up to isomorphisms) standard real field and the equality 
reads: $\mathbb{C}^2\simeq \mathbb{R}^4$. Instead, one can use an automorphism 
of $\mathbb{C}$ to obtain (non-canonical) isomorphism ${ ^{\ast}C}\overset{\rm 
iso}{\simeq}\mathbb{C}$ and thus $\mathbb{C}^2\simeq_{\rm iso}{ ^{\ast}R}^4$. 
Given different ${ ^{\ast}R}^4$'s one gets different automorphisms of 
$\mathbb{C}$ and thus different realizations of the isomorphism above. It 
follows that one can use different wild automorphisms of $\mathbb{C}$ to 
distinguish (index) different non-standard models of $\mathbb{R}$. Given 
$\mathbb{R}^4$ locally modified by ${\cal B}$ and taking its classical limit which factors through ${ ^{\ast}R}_1,{ ^{\ast}R}_2$, this results in the 
exotic $R^4_{1,2}$ and thus the correspondence follows:
\begin{corollary}
Pairs $(\alpha_1,\alpha_2)$ of automorphisms of $\mathbb{C}$, where at least 
one automorphism is wild, distinguishes different exotic $R^4_{1,2}$'s.
\end{corollary}
This relation can be expressed in terms of eq. (\ref{15}) which for $n=4$ and by 
turning to the $\mathbb{C}$ leads to the fully external description:
\begin{equation}\label{15-4-C}
\mathbb{C}^2\simeq \mathbb{R}^4\overset{2nd\to 1st}{\longrightarrow}{ 
^{\ast}R_1}^4\to { ^{\ast}C}_{1}\overset{({\rm iso},0)}{\to} \mathbb{C}\times 
\mathbb{C}\overset{(0,{\rm iso})}{\leftarrow} { ^{\ast}C}_{2}\leftarrow{ 
^{\ast}R_2}^4  \overset{2nd\to 1st}{\longleftarrow}\mathbb{R}^4\simeq 
\mathbb{C}^2.
\end{equation}
The middle $\mathbb{C}\times \mathbb{C}$ product emerges from the component-wise 
automorphisms of $\mathbb{C}$ giving rise to the isomorphisms $\alpha_i:{ 
^{\ast}C}_{i}\simeq \mathbb{C}, i=1,2$ and this is the pair 
$(\alpha_1,\alpha_2)$ which represents exotic $R^4_{1,2}$.
The relation is, however, highly non-constructive, similarly to the wild automorphisms of 
$\mathbb{C}$ and the ultrafilters constructions. 

As we observed the wild automorphisms of $\mathbb{C}$ should somehow allow for 
the assignment of finite values to some divergent expressions. We can make this 
point more tractable by turning to the relation with exotic $R^4_{1,2}$ and 
making use of the very special properties of ${\cal B}$. So we turn again to 
(\ref{15}) from (\ref{15-4-C}). The point is that ${\cal B}$ locally modifies 
$\mathbb{R}^4$ and the theory of distributions in ${\cal B}$ looks very special, 
namely all distributions in ${\cal B}$ are regular (constructive and w.r.t. the 
smooth real line and natural numbers) and each external distribution is 
canonically mapped into the internal one. In fact it holds (\cite{MR1991}, Th. 
3.6 p. 324 and Remark on p. 322):
\begin{theorem}[Moerdijk, Reyes, 1991]\label{Th4}
In ${\cal B}$ for every distribution $\mu$ on $R^n$ there exists a 
predistribution (function) $\mu_0:R^n\to R$ such that for all $f\in F_n$:
\[ \mu(f)=\int f(x)\mu_0(x) {\rm d}x .\]
\end{theorem}
Here $F_n$ denotes the internal space of test functions in dimension $n$. Also 
as stated by Theorem 3.15.3 p. 336 in \cite{MR1991}, there exists a bijection 
between the external distributions in Set and the internal in ${\cal B}$ given 
by the global section functor $\Gamma:{\cal B}\to {\rm Set}$.
In particular, the product and the square roots of distributions are thus 
well-defined in ${\cal B}$ as operations on the representing internal 
functions. 

\subsection{Renormalization in the coordinate space}
Now we can discuss the problem of renormalization in perturbative quantum field 
theory based on this special representation of distributions in ${\cal B}$. Note 
also that in ${\cal B}$ the standard NNO, i.e. $\mathbb{N}$, is replaced with the 
smooth NNO, $N_{\cal B}$, such that `finite' in ${\cal B}$ is `infinite' externally 
in Set. Thus indeed ${\cal B}$ is a natural category for 
addressing renormalization questions.
Given the interaction Lagrangian ${\cal L}=\frac{\lambda}{k!}\phi^k_I$ of the 
$\phi^k$ neutral scalar massive quantum field theory its $S$-matrix is 
determined in Dyson series representation, as \cite{Kreimer2004}:  
\begin{equation}\label{S1} S=\sum_{n=0}^{\infty}\frac{i^n}{n!}\int_{M^n}T({\cal 
L}_I(x_1){\cal L}_I(x_2)...{\cal L}_I(x_n)){\rm d}x_1{\rm d}x_2...{\rm 
d}x_n\end{equation}
where $M=\mathbb{R}^{1,3}$, $M^n=M\times ...\times M$ $n$-times and $T$ stays 
for the time-ordered products of the operator-valued distributions ${\cal L}_I$, 
hence $S$ is the operator-valued distribution either. The time ordering is defined 
for two operator valued functions $A,B$ on $M$ as (we follow the presentation in 
\cite{Kreimer2004}):
\begin{equation}\label{20}
T(A(x_1)B(x_2))=\Theta(x_1^0-x_2^0)A(x_1)B(x_2)+\Theta(x_2^0-x_1^0)B(x_2)A(x_1)
\end{equation}
where $\Theta(x)$ is the Heaviside function on $\mathbb{R}$, i.e. $\Theta(x)=0, 
x<0$ and $\Theta(x)=1, x\geq 0$. Here $x_i^0, i=1,2$ are time coordinates of 
$x_i\in M,i=1,2$. However, as noted in \cite{Kreimer2004} for general operator-valued irregular 
distributions one cannot create the products of them by 
discontinuous functions like $\Theta$. If one, however, works outside the thick 
diagonal in $M^n=M\times ...\times M$, $D_n=\{x\in M^n:\exists _{i\neq j}x_i=x_j 
\}$, then the $\Theta$ is continuous hence the product (\ref{20}) well-defined. 
This is the core of the various problems in perturbative QFT, let us quote the 
opinion of Authors of \cite{Kreimer2004}:
\begin{quotation}
In fact the mathematical origin for the appearance of short-distance 
singularities in perturbation theory is the ill-defined notion of time-ordering 
reviewed above. Epstein and Glaser
proposed a way to construct well-defined time ordered products $T_ n$ , one for 
each power
$n$ of the coupling constant, that satisfy a set of suitable conditions 
explained below, the
most prominent being that of locality or micro-causality. The power series $S$ 
constructed
by (\ref{S1}) using the Epstein-Glaser time-ordered product $T$ is a priori 
finite in every order, and
renormalization corresponds then to stepwise extension of distributions from 
$M_n\setminus D_n$ to $M_n$ . In general, distributions can not be extended 
uniquely onto diagonals. The resulting degrees of freedom are in one-to-one 
correspondence with the degrees of freedom (finite
renormalizations) in momentum space renormalization programs like BPHZ and 
dimensional regularization.
\end{quotation}
Instead of reviewing the Epstein-Glaser construction let us observe that for 
regular distributions the problem in (\ref{20}) does not arise since they can be 
represented by the operator-valued functions, their product is well-defined 
and they can be multiplied by $\Theta$. True problem arises for irregular 
distributions like Dirac $\delta$. Moreover, if all distributions were regular 
the Epstein-Glaser construction would give as the extension over $M_n\setminus 
D_n$ just the regular distributions we started with.

Now recall that in ${\cal B}$: 1. \emph{every} distribution is regular (Theorem 
\ref{Th4}), 2. \emph{every} distribution in Set can be naturally mapped to a 
distribution in ${\cal B}$ (Theorem 3.15.3 p. 336, \cite{MR1991}), and 3. the 
function $ ^{\cal B}\Theta :R\to R$ is continuous.
This observations motivate the following procedure:

Given $M^n$ ($n$-product of the Minkowski spacetime, $n=1,2,3...$) let the 
diagonal $D_n\subset U_{\beta}\in {\cal O}$ for some regular open cover 
$\{U_{\alpha}\}_{\alpha \in I}$ where $U_{\beta}\in \{U_{\alpha}\}_{\alpha \in 
I}$. Then, one locally modifies $M^n$ by ${\cal B}$ such that $U_{\beta}\in 
{\cal B}$ according to Def. \ref{defB}.

Under the procedure above one indeed has well-defined extensions of 
distributions over the diagonals $D_n$ in a sense of internal logic of ${\cal 
B}$ localized on $M^n$. Observe also that the local modification of $M^n$ 
implies some local modification of spacetime $M$ itself (if not, all factors in 
$M^n$ are not modified, hence $M^n$ neither).
\begin{corollary}
Varying the underlying geometry of a spacetime manifold by the local 
modification of its smooth structure by ${\cal B}$, i.e. $ ^{(\cal B)}M$, gives rise to 
the renormalization of some perturbative QFT when formulated on such modified 
manifolds.
\end{corollary}
Let us introduce the following additional suppositions: 

All the local deformations of $M$ are generated by the underlying local 
deformations by ${\cal B}$ of $\mathbb{R}^4$, and let the classical limit of 
them  factorize through some ${ ^{\ast}R}_{i}, i=1,2$, thus leading to exotic 
$R^4_{1,2}$. Then it follows:
\begin{corollary}\label{Th5}
The renormalization problem of some perturbative QFT can be translated into the geometry of some (Euclidean) exotic $R^4$ background which complements the Minkowski flat spacetime.  
\end{corollary}
One can restate the corollary as: \emph{Ultraviolet (UV) divergencies in some perturbative QFT determine exotic smoothness of the Euclidean $\mathbb{R}^4$ background}. We expect that ultraviolet divergencies counterterms of some perturbative QFT's on Minkowski spacetime are expressible in terms of the Riemannian (sectional) curvature of $R^4_{1,2}$. This Euclidean curved 4-background complements the Minkowski's one. Recall that exotic $R^4$'s are just Riemannian smooth 4-manifolds which can not 
be flat. Thus the Corollary \ref{Th5} indicates that a curvature in spacetime, hence 
nonzero density of gravitational energy emerges, when 
renormalization problem is solved geometrically. This connection with 
gravity is a rather universal, non-perturbative phenomenon of different perturbative QFT's and it is an important feature of the approach. 
 
\subsection{QM on smooth $R^4$ and model theory} 
The specific model-theoretic approach to exotic smoothness of open 4-manifolds 
like $\mathbb{R}^4$ presented here has also the advantage that one can still 
think in terms of local differentiation and (global) functions and arrive at the 
model-theoretic set-up. This is complementary to the approach via Riemannian 
structures and curvature, which anyway indicates that exotic $R^4$'s are 
`normal' smooth 4-manifolds and functions are local objects on them. We follow the work \cite{Krol2004b} and the case of 
exotic $R^4$'s is again crucial here. We work in the complementary picture and the analysis is based on model-theoretic 
tools but it is worth mentioning that strong connection of small exotic smooth $R^4$'s with QM formalism, noncommutative spaces, QFT and quantum 
gravity, was indeed shown and developed by purely geometric and topological methods 
(see e.g. \cite{TAMJK2011,TAMJK2014,TAMJK2010}).  
\begin{lemma}
Let $R^4$ be some exotic smoothness structure on $\mathbb{R}^4$. There has to 
exist a continuous (non-standard smooth) real-valued function on $\mathbb{R}^4$ 
which would be smooth on $R^4$, or a continuous real-valued function smooth on $\mathbb{R}^4$ but merely continuous on $R^4$.   
\end{lemma}
If such function did not exist that means that the precisely the same functions would be smooth in 
both structures and the smoothness structures would be equivalent 
and manifolds diffeomorphic (being homeomorphic). $\square$

So, let $f:\mathbb{R}^4\to \mathbb{R}, f\in C^0(\mathbb{R}^4)$ and $f\in 
C^{\infty}({R^4})$ so $f$ is exotic smooth. $f$ can not be everywhere standardly 
differentiable on $\mathbb{R}^4$ but, when changing the smoothness structure 
into $R^4$, it can. Moreover, the differentiation is locally the same as a
standard one, since $R^4$ is a Riemannian smooth 4-manifold. What would happen if one 
tried to differentiate globally any nondifferentiable continuous function? One should follow the pattern of generalized differentiation of functions or distributions. Outside the domains where the function is not continuous, the differentiation agrees with normal local differentiation. We are looking for the model-theoretic compensation (representation) for such global `non-standard' distributional differentiation. The result is precisely the $\mathbb{R}^4$ locally modified by ${\cal B}$.

Namely, it is always possible to choose open neighborhoods containing 
the nonsmooth domains of the function $f$ such that in these domains the 
functions would be represented by \emph{regular} distributions. However, iterating
differentiation of them leads to irregular distributions as well, like Dirac 
$\delta$-distribution. Then, we can turn to a $\mathbb{R}^4$ locally modified by 
${\cal B}$ such that the neighborhoods are internal in ${\cal B}$ and every external distribution, also irregular, is represented internally by regular one (Theorem 3.15.3 p. 336, \cite{MR1991}), i.e. by some 
internal \emph{smooth} function. This is the model-theoretic smoothing of continuous functions on $\mathbb{R}^4$. Taking the classical nontrivial limit of this local modification by ${\cal B}$ the result is some  exotic $R^4$ as in Theorem \ref{th2}. On the contrary, every local modification by ${\cal B}$ sends some irregular distributions to the internal smooth functions. Thus the following definition is natural and direct in this context: we call the modification by ${\cal B}$ the \emph{model-theoretic representation of an exotic smooth structure on $\mathbb{R}^4$} \cite{Krol2004b} provided it sends some irregular distributions to the internal smooth functions. Let exotic smooth $R^4_{1,2}$ be the classical limit of our $ ^{(\cal B)}R^4$ which factorizes through ${ 
^{\ast}R}_i,i=1,2$.
\begin{lemma}
Let the model-theoretic representation of the exotic smooth $R^4_{1,2}$ be $ ^{(\cal B)}R^4$. In the classical trivial, i.e. standard $\mathbb{R}^4$
limit, the space of exotic smooth functions on $R^4_{1,2}$ contains some irregular external distributions
 on the standard $\mathbb{R}^4$.
\end{lemma}
First, in the classical limit we do not have the dependence on ${\cal B}$ any 
longer.
Next, suppose that classical limit as in the formulation of the lemma does not contain the 
distribution. Then the global differentiation of every smooth function on 
$R^4_{1,2}$ agrees with the global standard differentiation on $\mathbb{R}^4$. 
So, the smoothness structure of $R^4_{1,2}$ has to be the standard one. $\square$

Next consider the Fourier transform of smooth functions ${\rm FT}: C^{\infty}(\mathbb{R}^4)\to C^{\infty}(\mathbb{R}^4)$. Let us represent the discontinuous functions in some open neighborhood by the corresponding irregular distributions as before. FT extends over the space of $L^2$-functions and distributions on $\mathbb{R}^4$ thus over $C^{\infty}(R^4_{1,2})$ in the $\mathbb{R}^4$ representation. The image of such Fourier operator is again $C^{\infty}(R^4_{1,2})$. This is the core of the interpretation of QM formalism on exotic $R^4$.
\begin{lemma}
The FT of $\delta$ and $\delta$-distribution itself, they both belong to the 
model-theoretic representation of an exotic smooth $R^4$ in the standard $\mathbb{R}^4$ limit.
\end{lemma}
We would like to interpret this result directly on exotic $R^4$. Note that the FT of $\delta$ is $\sim 1$ and it is geometrically a straight 
line, say coordinate axes, in the standard structure. However, this line can not be any smooth coordinate line in any exotic $R^4$, since this would give the factorization and the collapse of the structure to the standard one. However, the tangent space of every exotic $R^4$ is trivial, i.e. 
$TR^4\simeq T\mathbb{R}^4\simeq T_0\mathbb{R}^4$ ($R^4$ is contractible) and we 
consider this $1(x)$ as the coordinate line in the tangent space $TR^4$ 
\cite{Krol2004b}.\footnote{One could also think about such structures as having 
the generalized tangent spaces like e.g.  $TR^4\oplus { ^{\ast}T}R^4$. Indeed, 
one can relate \cite{TAMJK2009} some small exotic $R^4$'s with deformations of Hitchin 
structures (gerbes) defined on $TS^3\oplus { ^{\ast}T}S^3, S^3\subset 
\mathbb{R}^4$.} This coordinate line is spanned by $\sim 
\partial_x$ in the generator tangent space.  Thus FT mixes the 
standard tangent space with coordinate space $R^4$ and thus 
$\partial_x$ is sent to the multiplication operation in the 
model-theoretic representation of exotic $R^4$.
Given a large exotic $R^4$ (which can not be embedded into the standard $\mathbb{R}^4$) its contraction to a ball in $\mathbb{R}^4$ gives rise to:
\begin{theorem}[Corollary 4, \cite{Krol2004b}]
One can interpret the noncommutative relations of the position and momentum 
operators in the, contracted to a 4-ball, classical limit of the model-theoretic 
representation of a large exotic smooth ${R}^4$.\end{theorem}
Based on this interpretation the mechanism of decoherence in 
spacetime was proposed where QM effects disappear by taking uncontracted limit of such contracted $R^4$ \cite{Krol2004b}.

\section*{Acknowledgement}
The author appreciates much the important and fruitful discussions with Torsten Asselmeyer-Maluga and Krzysztof Bielas within the years about the wide range of topics appearing in the Chapter. 

%
%\addcontentsline{toc}{section}{Appendix}

\section*{Appendix}\label{App}
\subsection*{Weak arithmetic in smooth toposes}
In order to work constructively in arbitrary topos the correct logic is intuitionistic - one avoids 
the axiom of choice (AC) and the law of excluded middle (e.g. \cite{Moerdijk1992}). Next, instead of the axiom of choice, and even 
finite AC, one has the axiom of bounded search \cite{MR1991} as in (\ref{3}) 
below, recursion rule is replaced by the finitely presented type recursion 
(\ref{2}) and full induction is replaced by the following (\ref{1}) coherent 
induction scheme:\footnote{These axioms are written within the varying types 
formalism of S. Feferman \cite{MR1991,Fef1985}}

\begin{eqnarray}\label{1}
{\rm Ind}:\;\phi(0) \wedge \forall_{x\in N}(\phi(x)\to 
\phi(x+1))\to\forall_{x\in N}\phi(x),\; {\rm  for}\; \phi \;{{\rm 
coherent}}\\\label{2}
{\rm Rec}:\;\forall_{f\in S^{S\times T}}\forall_{a\in S^T}\exists !_{g\in 
S^{N\times T}}\forall_{x\in T}(g(0,x))=\\\nonumber
a(x)\wedge \forall_{n\in N} g(n+1,x)=f(g(n,x))\\\label{3}
{\rm weak\,AC}:\;\forall_{A\in P(N\times N)}(\forall_{n\in N}\exists_{m\in 
N}A(n,m)\to \\\nonumber
\forall_{n_0\in N} \exists_{m_0\in N}\forall_n n\leq n_0 \exists_{m}m\leq m_0 
\wedge A(n,m)).
\end{eqnarray}
The type $S$ in (\ref{2}) has to be finitely presented and the formula $\phi$ in 
(\ref{1}) coherent (e.g. \cite{MR1991} pp. 297-298) which results in further 
weakening of the logic.\footnote{$P(N\times N)$ is the power set of $N\times N$ 
and $A(n,m)$ means $(n,m)\in A\in P(N\times N)$.}  
Given such substantial weakening of the logic and arithmetic one gains the 
degree of indistinguishability of the standard and certain non-standard models of 
natural numbers. 
These weak properties are augmented by the usual subset of PA axioms (still in 
the intuitionistic logic):  
\begin{eqnarray}\label{4}
N{\rm \; is\; a\; subtype\; of\;} R;\\\label{5}
R {\rm \;is\;Archimedean}: \forall_{x\in R}\exists _{n\in N} z<n;\\\label{6}
0\in N\, {\rm and}\, \forall_{x\in R}(x\in N\to  x+1 \in N)\, {\rm and}\, 
\\\nonumber
\forall_{x\in R}(x\in N \wedge x+1=0 \to \bot).
\end{eqnarray}
As shown by Moerdijk and Reyes \cite{MR1991} these properties characterizing 
weak intuitionistic arithmetic along with coherent formulas and type 
restrictions as in (\ref{1},\ref{2}) above, are fulfilled in some \emph{smooth} 
toposes like smooth Zariski topos ${\cal Z}$ or Basel topos ${\cal B}$.

\addcontentsline{toc}{section}{Appendix}


\begin{thebibliography}{xxxx}
\bibitem{Benioff2015} P. Benioff, ``Space and time dependent scaling of numbers in mathematical structures: Effects on physical and geometric quantities'', will appear in Quantum Information Processing, Howard Brandt memorial issue.
\end{thebibliography}
\end{document}